\documentclass{appolb}
\usepackage{graphicx}
\usepackage{amsmath,amsfonts,amssymb,amscd,amsxtra,amsthm}
\usepackage[dvipsnames]{xcolor} 

\begin{document}
\title{Pentaquarks and Maxim V. Polyakov%
}
\author{Hyun-Chul Kim~\footnote{hchkim@inha.ac.kr}
\address{Hadron Theory Group and Department of Physics, Inha
  University, 100 Inha-Ro, Incheon 22212, Korea}
}
\maketitle
\begin{abstract}
This brief review is dedicated to the memory of Maxim V. Polyakov and
his pioneering contributions to pentaquark physics. We focus on his
seminal 1997 work with Diakonov and Petrov that predicted the
$\Theta^+$ pentaquark, a breakthrough that initiated an intense period
of research in hadron physics. The field faced a significant setback
when the CLAS Collaboration at Jefferson Lab reported null results in
2006, leading to a dramatic decline in light pentaquark
research. Nevertheless, Maxim maintained his scientific conviction,
supported by continued positive signals from DIANA and LEPS
collaborations. Through recent experimental findings on the $\Theta^+$
and the nucleon-like resonance $N^*(1685)$, we examine how Polyakov's  
theoretical insights, particularly the prediction of a narrow width 
($\Gamma \approx 0.5$-$1.0$ MeV), remain relevant to our understanding 
of the $\Theta^+$ light pentaquark. 
\end{abstract}

\section{Remembering Maxim}

I first met Maxim Vladimirovich Polyakov in October 1993 at the Institute 
for Theoretical Physics II at Ruhr-University Bochum. He and Pasha 
Pobylitsa came as guest scientists, both fresh-baked Ph.D.'s like myself. 
After a couple of discussions, I was surprised by their maturity in 
theoretical physics. They were not fresh-baked theoretical physicists but 
full-fledged ones. This was no surprise, as their advisor Mitya Diakonov 
once told me that he accepted only students in his group, who could
challenge him intellectually. Rather than feeling discouraged by their
expertise, I decided to learn from them. Our discussions were
characterized by direct, unvarnished feedback - ``\emph{Hyun-Chul,
  you're absolutely wrong!}'' was a common refrain. Maxim told me that
such a discussion must be natural in physics, and was a long tradition
in the Landau and Gribov school. 

I worked with Maxim on various subjects. Of all the research I
conducted with him, the study on nucleon tensor charges stands out as
the most memorable~\cite{Kim:1995bq}. I received an email from Maxim
in Autumn 1995, while he was visiting the University of Bern. In the
email, he wrote: 
``\emph{My goodness, Hyun-Chul. I totally forgot the tensor charges of
  the nucleon! We can compute them in our model. Since you have
  already calculated the axial-vector form factors of the nucleon, you
  can immediately derive the tensor charges of the nucleon. In this
  case, anomalous contributions come from the real part of the
  effective chiral action.}'' After I replied, I began computing the
nucleon tensor charges, which took about ten days. Once he returned to
Bochum, we discussed the results and physical implications of the
nucleon tensor charges. We wrote a manuscript together. It was truly
an amusing experience. 

When I got a permanent position in Korea, I regularly visited
Bochum. One of the main reasons for my visits was to discuss physics
with Maxim. In doing theoretical physics, one of the most important
things is to find the right person for discussion. Maxim was such a
person. When I had an idea, I always wanted to discuss it with him to
see if it was viable. He also visited Korea several times and even
attended my student's wedding ceremony. He enjoyed visiting Korea
because the atmosphere reminded him of Siberian mentality and
hospitality.

\begin{figure}[htp]
  \centering
\includegraphics[scale=1.4]{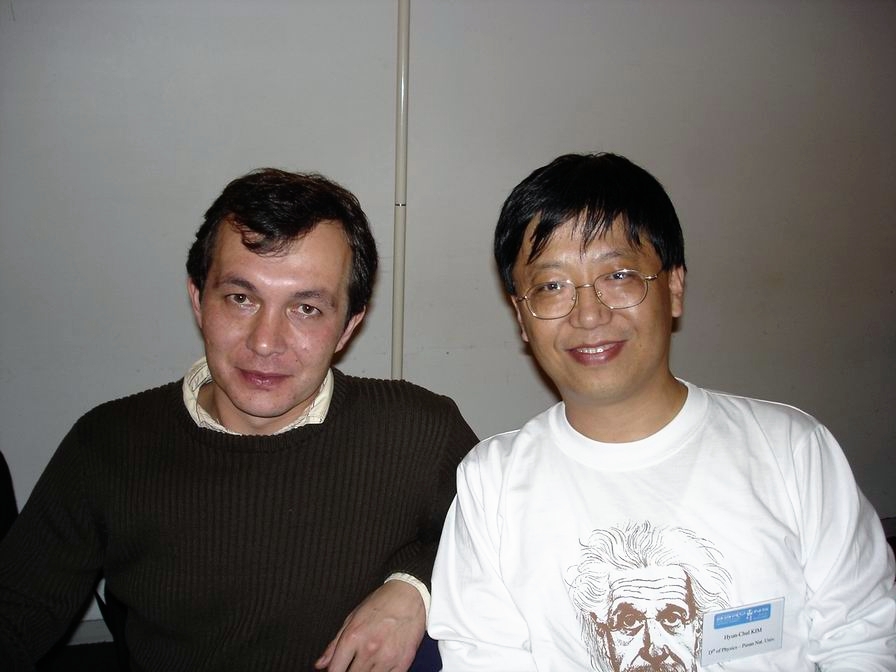}  
  \caption{Maxim V. Polyakov and Hyun-Chul Kim at the 10th
    International Conference on Structure of Baryons (Baryons 2004) in
    Paris, 2004.} 
  \label{fig:1}
\end{figure}
That's how we became colleagues, then friends, and finally brothers
when he gave me a small silver cross, saying, ``\emph{In Russia, giving a
silver cross to somebody means that you are my brother.}'' His
contributions to physics were profound: the prediction
of the light pentaquark with Diakonov and
Petrov~\cite{Diakonov:1997mm}, pioneering 
work on hard exclusive electroproduction of two
pions~\cite{Polyakov:1998ze}, significant advances in generalized
parton distributions~\cite{Goeke:2001tz}, and fundamental insights
into the mechanical structure of the
proton~\cite{Polyakov:1999gs}. His 2002 work on proton mechanical 
structure and gravitational form factors~\cite{Polyakov:2002yz,
  Polyakov:2018zvc} remains particularly relevant for the upcoming
electron-ion collider (EIC) experiments aimed at understanding proton
mass and spin~\cite{Accardi:2012qut, AbdulKhalek:2021gbh,
  Burkert:2022hjz}. 

Maxim's sudden passing left an irreplaceable void, both personally and in 
our field. His ability to bridge sophisticated theory with experimental 
observables was rare and valuable. Beyond his scientific achievements, I 
lost a brother who shared not only physics discussions but also life's 
broader journey. His contributions continue to guide our understanding of 
hadronic physics, especially as we approach the era of new experimental 
facilities. 
\section{A very brief history on pentaquarks} 
In the present review, I will focus on the pentaquark states among
many important Maxim's works. This can be a complementary one to
recent reviews by Amaryan~\cite{Amaryan:2022iij},
Prasza{\l}owicz~\cite{Praszalowicz:2024zsy}, and
Strakovsky~\cite{Strakovsky:2024ppo}. The paper with the title
``Exotic anti-decuplet of baryons: 
prediction from chiral solitons''~\cite{Diakonov:1997mm} predicted the
mass and width of an $S = +1$ baryon to be $M_{\Theta^+} \approx 1530$
MeV and $\Gamma_{\Theta^+} \approx 15$ MeV, which is a member of the
baryon antidecuplet ($\overline{10}$). The quark content of the $S =
+1$ baryon is known to be $uudd\bar{s}$, that is, a pentaquark
state. This new baryon was at first called $Z^+$, and was later
christened $\Theta^+$ by Diakonov~\cite{Strakovsky:2024ppo}. The
existence of the pentaquark $\Theta^+$ was for the first time
experimentally confirmed by the LEPS Collaboration at
SPring-8~\cite{LEPS:2003wug}. Then many experimental 
groups~\cite{DIANA:2003uet, SAPHIR:2003lnh, CLAS:2003yuj,
  CLAS:2003wfm, HERMES:2003phe, SVD:2004uuh, ZEUS:2004lje,
  Asratyan:2003cb} consecutively confirmed its existence. The 
findings were phenomenal and triggered numerous works on the light
pentaquark $\Theta^+$. 

Pentaquark states were first investigated in $K^+N$ and $K^+d$
experiments in the 1960s~\cite{Cool:1966zz, Abrams:1967zza,
  Bugg:1968zz, Abrams:1969obq} as well as in photoproduction
experiments~\cite{Tyson:1967zz}. The experimental signals observed in
these studies were all broad and were hypothesized to be pentaquark
states belonging to either the baryon antidecuplet or the baryon
eikosiheptaplet ($\mathbf{27}$). Golowich analyzed the data on $K^+ N$
scattering, and proposed the existence of $Z_0^*(1700)$ and
$N^*(1750)$~\cite{Golowich:1971bg}. However, all these 
proposed pentaquark states had broad widths. 
The mass of the $\Theta^+$ was also predicted in Skyrme
models~\cite{Biedenharn:1984qg, Praszalowicz:1987}. Notably, while
Biedenharn et al.~\cite{Biedenharn:1984qg} mentioned that they
eliminated an unwanted antidecuplet
($M_{\overline{\mathbf{10}}}\approx 1500$ MeV), Prasza{\l}owicz
predicted the mass of the pentaquark with strangeness $S=+1$ as
$M_{\overline{\mathbf{10}}}\approx 1530$ MeV, stating that whether
this prediction was a success or a drawback of the model would depend
on the reader's perspective\cite{Praszalowicz:1987}. Therefore, the
significance of Ref.~\cite{Diakonov:1997mm} lies not only in its mass
prediction but also in its prediction of the small width of the
$\Theta^+$. 

High-energy experiments reported negative results for the existence of
the $\Theta^+$\cite{BES:2004kia, Pinkenburg:2004ux, Knoepfle:2004zth,
  SPHINX:2004lbx, Belle:2005mke, Litvintsev:2004yw, BaBar:2004phi}. In
2006, the CLAS Collaboration reported null results regarding the
existence of the $\Theta^+$\cite{CLAS:2006czw, CLAS:2005koo,
  CLAS:2006anj}. The E522 experiment at KEK also searched for the
$\Theta^+$ using the inclusive reactions $\pi^- p \to K^+ X$ and $K^+
p\to \pi^+ X$, but did not find any significant
signal~\cite{KEK-PSE522:2006xah, Miwa:2007xk}. Following these
results, publications on pentaquarks decreased dramatically, and the
scientific attitude toward light pentaquarks became increasingly
skeptical until the discovery of heavy pentaquark states
$P_{c\bar{c}}$\cite{LHCb:2015yax, LHCb:2019kea,
  LHCb:2021chn}. Nevertheless, the LEPS and DIANA Collaborations
continued to report evidence for the existence of the
$\Theta^+$\cite{LEPS:2008ghm, Nakano:2010zz, DIANA:2006ypd,
  DIANA:2009rzq, DIANA:2013mhv}. For detailed discussions of both
positive and negative experimental results, see a recent paper by
Amaryan~\cite{Amaryan:2022iij}. 

Several important questions regarding the pentaquark $\Theta^+$ remain
unanswered. Following the initial discovery by the LEPS Collaboration,
multiple experiments confirmed the existence of the $\Theta^+$. If the
$\Theta^+$ indeed does not exist, what were the peaks that multiple
experimental groups identified as the $\Theta^+$? Most experiments
that yielded positive results for the existence of the $\Theta^+$ used
real and virtual photons scattered off the nucleon, deuteron, or
nuclei. However, it is noteworthy that the null results from the CLAS
Collaboration also came from photoproduction experiments. In contrast,
the experiments that did not observe the $\Theta^+$ were primarily
based on $e^+e^-$ annihilation to hadrons. The reason for this
experimental discrepancy remains unclear. 

With the exception of the DIANA Collaboration~\cite{DIANA:2003uet}, no
experiments used kaon beams. The $\Theta^+$, if it exists, would
primarily decay into $K^+n$ or $K^0p$. This suggests that a kaon beam
with appropriate momentum could be used to form the $\Theta^+$
directly~\cite{Sibirtsev:2004bg, Sekihara:2019cot}. Notably, the DIANA
Collaboration has continued to report evidence for the existence of
the $\Theta^+$\cite{DIANA:2006ypd, DIANA:2009rzq,
  DIANA:2013mhv}. DIANA utilized a liquid xenon bubble chamber through
which the $K^+$ beam passed. Using this setup in the $K^+
\mathrm{Xe}\to K^0 p \mathrm{Xe}'$ reaction, DIANA reported the mass
and width of $\Theta^+$ as $M_{\Theta^+} =(1538\pm 2)$ MeV and
$\Gamma_{\Theta^+} = (0.34\pm 0.10)$ MeV, respectively. Additionally,
a break-away group from the CLAS Collaboration reported the
observation of the $\Theta^+$ in ${}^1\mathrm{H}(\gamma, K_S^0)X$
through interference with $\phi$-meson
production\cite{Amaryan:2011qc}. 

The use of kaon beams provides a significant advantage in searching
for $S=+1$ pentaquark baryons, as the $\Theta^+$ can be formed through
direct formation. Several experimental proposals to search for the
$\Theta^+$ have been put forward. Following the suggestion by Sekihara
et al.\cite{Sekihara:2019cot}, Ahn and Kim\cite{Ahn:2023JKPS} proposed
searching for the $\Theta^+$ using the $K^+d\to K^0 pp$ reaction. The
KLF Collaboration proposed investigating elastic $K_L p \to K_S p$ and
charge-exchange $K_Lp\to K^+n$ reactions~\cite{KLF:2020gai,
  Amaryan:2024koq}. These new proposals may definitively resolve the
question of the $\Theta^+$'s existence in the near future. 

In the work of Diakonov, Petrov, and Polyakov~\cite{Diakonov:1997mm},
the critical assumption was that the $P_{11}$ resonance $N^*(1710)$
was a nonstrange member of the baryon antidecuplet. Polyakov and
Rathke examined the radiative decays of this nonstrange member and
discovered that the transition magnetic moment for the neutron channel
is much larger than that for the proton, a phenomenon they termed the
\emph{neutron anomaly}~\cite{Polyakov:2003dx}. Based on this finding,
they proposed that this nonstrange member could be identified in
\emph{antidecuplet friendly photoproductions} such as $\gamma n\to K^+
\Sigma^-$, $\gamma n \to \eta n$, and $\gamma n\to (\pi\pi)_{I=1}
N$. The key insight here is that a neutron target is more favorable
for producing neutronlike pentaquarks. 

Kuznetsov et al. first reported the
existence of a nucleon-like narrow resonance at around $1.68$
GeV~\cite{GRAAL:2004ndn, GRAAL:2006gzl}, which was subsequently
confirmed by the CBELSA/TAPS Collaboration~\cite{CBELSA:2007vce,
  CBELSA:2008epm, Jaegle:2011sw}, A2 Collaboration~\cite{A2:2014pie},
and at Laboratory of Nuclear Science (LNS), Tohoku
University~\cite{Miyahara:2007zz}. This narrow resonance 
could naturally be identified as a neutronlike pentaquark belonging to
the baryon antidecuplet within the $\chi$QSM~\cite{Polyakov:2003dx,
  Kim:2005gz}. With this new pentaquark resonance, the experimental
data on $\gamma n \to \eta n$ were well described within reaction
models~\cite{Choi:2005ki, Choi:2007gy, Suh:2018yiu}. Furthermore, this
narrow $N^*(1685)$ resonance played an essential role in describing
the $\gamma n \to K^0 \Lambda$ reaction near the
threshold~\cite{Kim:2018qfu}. 
However, alternative interpretations were proposed, including:
coupled-channel effects of known nucleon
resonances~\cite{Shklyar:2006xw, Shyam:2008fr}, contributions from
intermediate strangeness states~\cite{Doring:2009qr}, and interference
in the partial wave between contributions from the well-known
$N^*(1535)$ and $N^*(1650)$ resonances~\cite{Anisovich:2015tla}. While
additional experimental evidence is needed to definitively identify
the nature of the narrow resonance $N^*(1685)$, Maxim argued for the
simplest interpretation of $N^*(1685)$ as a neutron-like pentaquark,
citing the \emph{principle of economy} or \emph{Occam's
  razor}~\cite{Boika:2014aha}. 

\section{Masses of $\Theta^+$ and $N^*(1685)$}
In this Section, we will briefly review an extended analysis from
Ref.\cite{Diakonov:1997mm}, based on Refs.\cite{Yang:2010fm,
Yang:2013tka}. 
Witten demonstrated that in the large $N_c$ limit of quantum
chromodynamics (QCD), where $N_c$ is the number of colors, a baryon
emerges as a bound state of $N_c$ valence quarks in a pion mean
field~\cite{Witten:1979kh}. The $N_c$ valence quarks generate an
effective pion mean field that arises from vacuum polarization, and
these same valence quarks are then bound by this self-consistently
generated field. This classical mean-field solution can also be
described as a chiral soliton with hedgehog symmetry, which represents
the minimal generalization of spherical symmetry~\cite{Pauli:1942kwa,
  Skyrme:1962vh}. 
The chiral quark-soliton model ($\chi$QSM)~\cite{Diakonov:1987ty,
  Christov:1995vm, Diakonov:1997sj} is founded on Witten's seminal
idea. Since meson fluctuations are suppressed in the large $N_c$
limit, the path integral over the pseudo-Nambu-Goldstone boson (pNGB)
fields can be evaluated using a saddle-point approximation. The chiral
soliton emerges as a solution by minimizing the classical nucleon mass
self-consistently, taking into account both the energies of the $N_c$
valence quarks (level quarks) and the sea quarks (Dirac continuum). 
A key strength of this pion mean-field approach is its ability to
describe both low-lying light baryons and singly heavy baryons within
a unified framework~\cite{Yang:2016qdz}. 

Since the classical nucleon represented by the chiral soliton lacks
quantum numbers such as spin and isospin, quantization is
necessary. While meson fluctuations are suppressed by the $1/N_c$
expansion, a complete consideration of zero modes, which arise from
rotational and translational symmetries, remains essential. This
zero-mode quantization in flavor SU(3) leads to the effective
collective Hamiltonian~\cite{Diakonov:1997mm, Blotz:1992pw,
  Yang:2010fm}: 
\begin{equation}
  \label{eq:1}
H = M_{\mathrm{cl}} + H_{\mathrm{rot}} + H_{\mathrm{sb}},  
\end{equation}
where $M_{\mathrm{cl}}$ corresponds to the classical mass of the
chiral soliton. $H_{\mathrm{rot}}$ represents the $1/N_c$ rotational
Hamiltonian given by: 
\begin{equation}
  \label{eq:2}
H_{\mathrm{rot}} = \frac1{2I_1} \sum_{i=1}^3 \hat{J}_i^2 +
\frac1{2I_2} \sum_{p=4}^7 \hat{J}_p^2,   
\end{equation}
where $J_i$ and $J_p$ are generators of the flavor SU(3) group, with
$J_i$ representing the usual spin operators. 
$I_{1,2}$ indicate the SU(3) soliton moments of inertia, determined
through the specific dynamics of chiral solitonic approaches such as
the $\chi$QSM or the Skyrme model. 
$H_{\mathrm{sb}}$ represents the explicit SU(3) symmetry breaking
term~\cite{Yang:2010fm}, which takes the form: 
\begin{eqnarray}
  \label{eq:3}
H_{\mathrm{sb}} &=& \left(m_{\mathrm{d}}-m_{\mathrm{u}}\right)
\left(\frac{\sqrt{3}}{2}\,\alpha\,
  D_{38}^{(8)}(\mathcal{R})
\;+\;\beta\,\hat{T_{3}}
\;+\;\frac{1}{2}\,\gamma\sum_{i=1}^{3}D_{3i}^{(8)}
(\mathcal{R})\,\hat{J}_{i}\right)\cr
 &  & +\;\left(m_{\mathrm{s}}-\overline{m}\right)\left(\alpha\,
   D_{88}^{(8)}(\mathcal{R})
\;+\;\beta\,\hat{Y}
\;+\;\frac{1}{\sqrt{3}}\,\gamma\sum_{i=1}^{3}D_{8i}^{(8)}
(\mathcal{R})\,\hat{J}_{i}\right)\cr
 &  &
      +\;\left(m_{\mathrm{u}}+m_{\mathrm{d}}+m_{\mathrm{s}}\right)\sigma,   
\end{eqnarray}
where $m_{\mathrm{u}}$, $m_{\mathrm{d}}$, and $m_{\mathrm{s}}$ denote
the current quark masses for up, down, and strange quarks,
respectively. Here, $\overline{m}$ indicates the average of up and
down quark masses. The $D_{ab}^{(\mathcal{R})}(\mathcal{R})$ desginate 
the SU(3) Wigner $D$ functions, while $\hat{Y}$ and $\hat{T_{3}}$ act
as operators for the hypercharge and isospin third component,
respectively. The parameters $\alpha$, $\beta$, and $\gamma$ can be
written in terms of the $\pi N$ sigma term, $\Sigma_{\pi N}$, and
soliton moments of inertia $I_{1,2}$ and $K_{1,2}$ as
\begin{equation}
\alpha
=-\left(\frac{2}{3}\frac{\Sigma_{\pi
      N}}{m_{\mathrm{u}}+m_{\mathrm{d}}}
-\frac{K_{2}}{I_{2}}\right),
\;\;\;\;\beta
=-\frac{K_{2}}{I_{2}},
\;\;\;\;\gamma
=2\left(\frac{K_{1}}{I_{1}}-\frac{K_{2}}{I_{2}}\right).
\label{eq:4}
\end{equation} 
The $\sigma$ is proportional to $\Sigma_{\pi N}$: 
\begin{equation}
  \label{eq:5}
\sigma\;\;=\;\;-(\alpha+\beta)
\;\;=\;\;\frac{2}{3}\frac{\Sigma_{\pi
    N}}{m_{\mathrm{u}}+m_{\mathrm{d}}}  . 
\end{equation}
The eighth of the generators of the SU(3) group is constrained by the
collective quantization:
\begin{equation}
  \label{eq:6}
J_{8}\;=\;-\frac{N_{c}}{2\sqrt{3}}B
\;=\;-\frac{\sqrt{3}}{2},\;\;\;\;
Y'\;=\;\frac{2}{\sqrt{3}}J_{8}
\;=\;-\frac{N_{c}}{3}\;=\;-1,  
\end{equation}
where $B$ denotes the baryon number. In the $\chi$QSM, this baryon
number emerges from the $N_c$ valence quarks occupying the discrete
level~\cite{Blotz:1992pw,Christov:1995vm}, whereas in the SU(3) Skyrme
model, it originates from the Wess-Zumino
term~\cite{Witten:1983tx,Guadagnini:1983uv,Jain:1984gp}. 
This constraint limits the allowed representations to $\mathrm{SU(3)}$
irreducible representations with zero triality. Consequently, the
permissible $\mathrm{SU(3)_{f}}$ multiplets include the baryon octet
($J=1/2$), decuplet ($J=3/2$), and antidecuplet ($J=1/2$), among
others. This feature underscores the success of collective
quantization and highlights a distinctive duality between a rigidly
rotating soliton and a constituent quark model. 

The baryon collective wave functions can be formulated using the SU(3)
Wigner $D$ functions in representation
$\mathcal{R}=\mathbf{8},\,\mathbf{10},\,
\overline{\mathbf{10}},\cdots$:   
\begin{eqnarray}
&&\langle A|\mathcal{R},\, B(Y\, T\, T_{3},\; Y^{\prime}\, J\,
J_{3})\rangle  \cr
& = & \Psi_{(\mathcal{R}^{*}\,;\, Y^{\prime}\, J\,
  J_{3})}^{(\mathcal{R\,};\, Y\, T\, T_{3})}(A)
  =  \sqrt{\textrm{dim}(\mathcal{R})}\,(-)^{J_{3}
+Y^{\prime}/2}\, D_{(Y,\, T,\, T_{3})(-Y^{\prime},\,
J,\,-J_{3})}^{(\mathcal{R})*}(A),  
\end{eqnarray}
where $Y,\, T,\, T_{3}$ correspond to the hypercharge, isospin,
and its third component for a given baryon, respectively. 
The symmetry-breaking term in the collective Hamiltonian
(Eq.(\ref{eq:3})) induces mixing between different $\mathrm{SU(3){f}}$
representations. Consequently, the collective wave functions
incorporate corrections from other allowed representations due to
SU(3) symmetry breaking 
\begin{eqnarray}
\left|B_{8}\right\rangle  
& = &\left|8_{1/2},B\right\rangle 
\;+\; c_{\overline{10}}^{B}\left|\overline{10}_{1/2},B\right\rangle 
\;+\; c_{27}^{B}\left|27_{1/2},B\right\rangle ,
\\
\left|B_{10}\right\rangle  
& =&\left|10_{3/2},B\right\rangle 
\;+\; a_{27}^{B}\left|27_{3/2},B\right\rangle 
\;+\;
a_{35}^{B}\left|35_{3/2},B\right\rangle ,
\cr
\left|B_{\overline{10}}\right\rangle  
& =&\left|\overline{10}_{1/2},B\right\rangle 
\;+\; d_{8}^{B}\left|8_{1/2},B\right\rangle 
\;+\; d_{27}^{B}\left|27_{1/2},B\right\rangle 
\;+\; d_{\overline{35}}^{B}\left|\overline{35}_{1/2},B\right\rangle.
     \nonumber 
\label{eq:8}  
\end{eqnarray}
The detailed expressions for the coefficients in Eq.~(\ref{eq:8}) are
available in Ref.~\cite{Yang:2010fm}. 

Beyond the explicit SU(3) symmetry breaking, we have incorporated
isospin symmetry breaking effects arising from the up and down current
quark mass difference, manifested in the first term of
$H_{\mathrm{sb}}$ in Eq.~(\ref{eq:3}). Furthermore, our analysis
includes isospin symmetry breaking from electromagnetic
self-interactions~\cite{Yang:2010id}. This comprehensive treatment of
symmetry breaking enables a more refined analysis compared to previous
studies~\cite{Diakonov:1997mm, Ellis:2004uz}. 

The zero-mode collective quantization with hedgehog symmetry functions
independently of any specific chiral solitonic approach. This
universality implies that under hedgehog symmetry and explicit flavor
SU(3) symmetry breaking, Eqs.~(\ref{eq:1}) and (\ref{eq:2}) take a
model-independent form. Consequently, rather than calculating
parameters within specific chiral soliton models, we can determine the
moments of inertia and other parameters ($\alpha$, $\beta$, $\gamma$,
etc.) using experimental and empirical data. 
To compute the baryon decuplet and antidecuplet masses, we utilize the
experimental baryon octet masses. Additionally, determining the
moments of inertia $I_1$ and $I_2$ requires at least two mass inputs
from the baryon antidecuplet. For this purpose, we adopt the masses of
$\Omega$ and $\Theta$: $M_{\Omega}=(1672.45\pm 0.29)$ MeV~\cite{PDG}
and $M_{\Theta^+}=(1524\pm 5)$ MeV~\cite{LEPS:2008ghm}. The inclusion
of isospin symmetry breaking allows for unambiguous determination of
all parameters. 

\begin{table}[htp]
 \centering \caption{Predicted masses of the baryon decuplet. The
   experimental data of decuplet baryons are taken from the Particle
   Data Group (PDG)~\cite{PDG}.} 
\begin{tabular}{cccccc}
\hline 
\multicolumn{2}{c}{Mass {[}MeV{]}} 
& $T_{3}$  
& $Y$  
& Experiment \cite{PDG}  
& Predictions
\tabularnewline
\hline 
$M_{\Delta}$  
& $\begin{array}{c}
\Delta^{++}\\
\Delta^{+}\\
\Delta^{0}\\
\Delta^{-}
\end{array}$  
& $\begin{array}{c}
\;\;3/2\\
\;\;1/2\\
-1/2\\
-3/2
\end{array}$  
& $\;\;1$  
& $1231-1233$  
& $\begin{array}{c}
1248.54\pm3.39\\
1249.36\pm3.37\\
1251.53\pm3.38\\
1255.08\pm3.37
\end{array}$\tabularnewline
\hline 
$M_{\Sigma^{\ast}}$  
& $\begin{array}{c}
\Sigma^{\ast+}\\
\Sigma^{\ast0}\\
\Sigma^{\ast-}
\end{array}$  
& $\begin{array}{c}
\;\;1\\
\;\;0\\
-1
\end{array}$  
& $\;\;0$  
& $\begin{array}{c}
1382.83\pm0.34\;\\
1383.7\pm1.0\\
1387.2\pm0.5\,
\end{array}$  
& $\begin{array}{c}
1388.48\pm0.34\\
1390.66\pm0.37\\
1394.20\pm0.34
\end{array}$
\tabularnewline
\hline 
$M_{\Xi^{\ast0}}$  
& $\begin{array}{c}
\Xi^{\ast0}\\
\Xi^{\ast-}
\end{array}$  
& $\begin{array}{c}
\;\;1/2\\
-1/2
\end{array}$  
& $-1$  
& $\begin{array}{c}
1531.80\pm0.32\\
1535.0\;\pm0.6\;\,
\end{array}$  
& $\begin{array}{c}
1529.78\pm3.38\\
1533.33\pm3.37
\end{array}$
\tabularnewline
\hline 
$M_{\Omega^{-}}^{\star}$  
& $\Omega^{-}$  
& $0$  & $-2$  
& $1672.45\pm0.29$  
& Input
\tabularnewline
\hline 
\end{tabular}
\label{tab:1} 
\end{table}
To validate our theoretical framework, we initially calculated the
masses of the baryon decuplet ($\mathbf{10}$). As shown in
Table~\ref{tab:1}, the pion mean-field approach yields results in
excellent agreement with the PDG data~\cite{PDG}, with the exception
of the $\Delta$ isobar masses. The discrepancy in $\Delta$ baryon
masses, ranging from 10 to 20 MeV, remains acceptable given their
substantial decay widths ($\Gamma_{\Delta}\approx (114-120)$
MeV). Having confirmed the reliability of the theoretical framework,
we can proceed with evaluating the baryon antidecuplet masses.  
\begin{table}[htp]
 \centering \caption{Comparison of the results for the masses of the
   baryon antidecuplet. 
 It is important to emphasize that the data from the NA49
 experiment~\cite{NA49:2003fxh} still requires independent experimental
 verification.}  
{\scriptsize 
\begin{tabular}{cccccc}
\hline 
\multicolumn{2}{c}{Mass} 
& Diakonov et al.~\cite{Diakonov:1997mm}
& $\chi\mathrm{QSM}$~\cite{Ledwig:2008rw}
& Exp.
& Ref.~\cite{Yang:2010fm} 
\tabularnewline
\hline 
$M_{\Theta^{+}}$  
& $\Theta^{+}$  
& $1530$   
& $1538$  
& $1524\pm 5$\cite{LEPS:2008ghm}
& Input
\tabularnewline
\hline 
$M_{N^{\ast}}$  
& $\begin{array}{c}
p^*\\
n^* \\
\end{array}$
& $1710^\star$
& $1653$
& ${\displaystyle 1686\pm 12}$
\cite{Kuznetsov:2008ii}
& $\begin{array}{c}
1688.18\pm10.53\\
1692.16\pm10.53
\end{array}$\tabularnewline
\hline 
$M_{\Sigma_{\overline{10}}}$  
& $\begin{array}{c}
\Sigma_{\overline{10}}^{+}\\
\Sigma_{\overline{10}}^{0}\\
\Sigma_{\overline{10}}^{-}
\end{array}$  
& $1890$  
& $1768$  
&  
& $\begin{array}{c}
1852.35\pm10.00\\
1856.33\pm10.00\\
1858.95\pm10.00
\end{array}$\tabularnewline
\hline 
$M_{\Xi_{3/2}}$  
& $\begin{array}{c}
\Xi_{3/2}^{+}\\
\Xi_{3/2}^{0}\\
\Xi_{3/2}^{-}\\
\Xi_{3/2}^{--}
\end{array}$  
& $2070$  
& $1883$  
& $1862\pm2$~\cite{NA49:2003fxh} 
& $\begin{array}{c}
2016.53\pm10.53\\
2020.51\pm10.53\\
2023.12\pm10.53\\
2024.37\pm10.53
\end{array}$\tabularnewline
\hline 
\end{tabular}
}
\label{tab:2} 
\end{table}
Table~\ref{tab:2} presents our calculated masses for the baryon
antidecuplet alongside results from previous
studies~\cite{Diakonov:1997mm, Ledwig:2008rw}. While Ledwig et
al.\cite{Ledwig:2008rw} derived the dynamical parameters within the
SU(3) $\chi$QSM framework under isospin symmetry constraints, the
current approach yields $N^*(1685)$ masses that are in good agreement
with experimental observations~\cite{Kuznetsov:2008ii}. Although the 
predicted $\Xi_{3/2}$ masses exceed those reported by the NA49
collaboration~\cite{NA49:2003fxh}, it should be noted that these
experimental values currently lack independent confirmation from other
experiments. 

Let us examine how the $N^*(1685)$ mass varies when the $\Theta^+$ mass
shifts from the LEPS data to the DIANA measurement ($M_{\Theta^+}=
(1538\pm 2)$ MeV). To simplify this analysis, we disregard isospin
symmetry breaking effects. Under these conditions, the $\Theta^+$ and
$N^*$ masses take the form: 
\begin{eqnarray}
  \label{eq:9}
M_{\Theta^+} &=& \overline{M}_{\overline{10}} - 2m_{\mathrm{s}} \delta,\cr  
M_{N^*} &=& \overline{M}_{\overline{10}} - m_{\mathrm{s}} \delta,
\end{eqnarray}
where $\overline{M}_{\overline{10}}$ denotes the mass-splitting center
of the baryon antidecuplet, determined by $\overline{M}_8 + 3/2I_2$
(see Ref.~\cite{Yang:2010fm} for details). The parameter $\delta$,
defined as $\delta = -\alpha/8 -\beta +\gamma/16$, governs the mass
splitting between members of the baryon
antidecuplet~\cite{Yang:2013tka}. 

Tables~\ref{tab:3} and \ref{tab:4} summarize the numerical values of
dynamical parameters. The parameter $c_{\overline{10}}$ represents the
octet-antidecuplet mixing amplitude, corresponding to
$c_{\overline{10}}^B$ in Eq.(\ref{eq:8}). Unlike previous
studies~\cite{Diakonov:1997mm, Ellis:2004uz} which used $\Sigma_{\pi
  N}$ as input, the present framework successfully predicts its value,
notably yielding a relatively small magnitude. The fourth column
displays results from the $\chi$QSM with all computed parameters. 
A comparison between Tables~\ref{tab:3} and \ref{tab:4} reveals
minimal variation in parameter values. The corresponding $N^*$ masses
are $(1690\pm 11)$ MeV and $(1701\pm 5)$ MeV, respectively, falling
between the predictions of Refs.~\cite{Ellis:2004uz} and
\cite{Diakonov:1997mm}. These findings suggest that the experimentally
observed $N^*(1685)$ mass~\cite{Nakano:2010zz} supports the $\Theta^+$
mass measurement from the LEPS collaboration. 

\begin{table}[htp]
\caption{The comparison of the dynamical parameters with those of
  Refs~\cite{Diakonov:1997mm, Ellis:2004uz, Ledwig:2008rw}. The masses 
  of the baryon antidecuplet members used as input are listed in the
  second row. The mass of the $\Theta^+$ is taken from the LEPS
  data~\cite{Nakano:2010zz}, i.e., $(1524\pm 5)$ MeV.} 
{\footnotesize 
\begin{tabular}{c||ccccc}
\hline 
\multicolumn{2}{c}{} 
& Diakonov et al.~\cite{Diakonov:1997mm} 
& \multicolumn{1}{c}{Ellis et al.~\cite{Ellis:2004uz}} 
& $\chi\mathrm{QSM}$\cite{Ledwig:2008rw} 
&  Present work I 
\tabularnewline
\hline 
\multicolumn{2}{c}{Input} 
& $N^{\ast}(1710\;\mathrm{MeV}$) 
& $\Theta^{+}(1539\pm2\;\mathrm{MeV})$
& $\cdots$ 
& $\Theta^{+}(1524\pm5\;\mathrm{MeV})$ 
\tabularnewline
\multicolumn{2}{c}{masses} 
&  
& $\Xi_{3/2}^{--}(1862\pm2\;\mathrm{MeV})$ 
& $\cdots$ 
&  
\tabularnewline
\hline 
\multicolumn{2}{c}{$\Sigma_{\pi N}$} 
& $45\;\mathrm{MeV}^{\star}$ 
& $73\;\mathrm{MeV}^{\star}$ 
& $41\;\mathrm{MeV}$ 
& $36.4\pm3.9\;\mathrm{MeV}$ 
\tabularnewline
\hline 
\multicolumn{2}{c}{$I_{1}^{-1}$} 
& $152.97$ $\mathrm{MeV}$ 
& $155.38$ $\mathrm{MeV}$ 
& $186.16$ $\mathrm{MeV}$ 
& $160.43\pm0.26$ $\mathrm{MeV}$ 
\tabularnewline
\multicolumn{2}{c}{$I_{2}^{-1}$} 
& $493.32\;\mathrm{MeV}$ 
& $402.71\;\mathrm{MeV}$ 
& $411.10\;\mathrm{MeV}$ 
& $469.83\pm6.71\;\mathrm{MeV}$ 
\tabularnewline
\multicolumn{2}{c}{$m_{s}\alpha$} 
& $-218\;\mathrm{MeV}$ 
& $-605\;\mathrm{MeV}$ 
& $-197\;\mathrm{MeV}$ 
& $-262.9\pm5.9\;\mathrm{MeV}$ 
\tabularnewline
\multicolumn{2}{c}{$m_{s}\beta$} 
& $-156\;\mathrm{MeV}$ 
& $-23\;\mathrm{MeV}$ 
& $-94\;\mathrm{MeV}$ 
& $-144.3\pm3.2\;\mathrm{MeV}$ 
\tabularnewline
\multicolumn{2}{c}{$m_{s}\gamma$} 
& $-107\;\mathrm{MeV}$ 
& $152\;\mathrm{MeV}$ 
& $-53\;\mathrm{MeV}$ 
& $-104.2\pm2.4\;\mathrm{MeV}$ 
\tabularnewline
\multicolumn{2}{c}{$c_{\overline{10}}$} 
& $0.084$ 
& $0.088$ 
& $0.037$ 
& $0.0434\pm0.0006$ 
\tabularnewline
\hline 
\multicolumn{2}{c}{$N_{\overline{10}}$} 
&  &  &  
& $1690\pm11\;{\rm MeV}$ 
\tabularnewline
\hline 
\end{tabular}
}
\label{tab:3}
\end{table}
\begin{table}[htp]
\caption{The comparison of the dynamical parameters with those of
  Refs~\cite{Diakonov:1997mm, Ellis:2004uz, Ledwig:2008rw}. The masses  
  of the baryon antidecuplet members used as input are listed in the
  second row. The mass of the $\Theta^+$ is taken from the DIANA
  data~\cite{DIANA:2013mhv}, i.e., $(1538\pm 2)$ MeV.}  
{\footnotesize 
\begin{tabular}{c||ccccc}
\hline 
\multicolumn{2}{c}{} 
& Diakonov et al.~\cite{Diakonov:1997mm} 
& \multicolumn{1}{c}{Ellis et al.~\cite{Ellis:2004uz}} 
& $\chi\mathrm{QSM}$\cite{Ledwig:2008rw} 
& Present work II
\tabularnewline
\hline 
\multicolumn{2}{c}{Input} 
& $N^{\ast}(1710\;\mathrm{MeV})$ 
& $\Theta^{+}(1539\pm2\;\mathrm{MeV})$
& $\cdots$ 
& $\Theta^{+}(1538\pm2\;\mathrm{MeV})$
\tabularnewline
\multicolumn{2}{c}{masses} 
&  
& $\Xi_{3/2}^{--}(1862\pm2\;\mathrm{MeV})$ 
& $\cdots$ 
&
\tabularnewline
\hline 
\multicolumn{2}{c}{$\Sigma_{\pi N}$} 
& $45\;\mathrm{MeV}^{\star}$ 
& $73\;\mathrm{MeV}^{\star}$ 
& $41\;\mathrm{MeV}$ 
& $37.5\pm1.1\;\mathrm{MeV}$
\tabularnewline
\hline 
\multicolumn{2}{c}{$I_{1}^{-1}$} 
& $152.97$ ${\rm MeV}$ 
& $155.38$ ${\rm MeV}$ 
& $186.16$ ${\rm MeV}$ 
& $160.43\pm0.26$ ${\rm MeV}$
\tabularnewline
\multicolumn{2}{c}{$I_{2}^{-1}$} 
& $493.32\;\mathrm{MeV}$ 
& $402.71\;\mathrm{MeV}$ 
& $411.10\;\mathrm{MeV}$ 
& $475.49\pm3.44\;\mathrm{MeV}$
\tabularnewline
\multicolumn{2}{c}{$m_{s}\alpha$} 
& $-218\;\mathrm{MeV}$ 
& $-605\;\mathrm{MeV}$ 
& $-197\;\mathrm{MeV}$ 
& $-281.5\pm6.30\;\mathrm{MeV}$
\tabularnewline
\multicolumn{2}{c}{$m_{s}\beta$} 
& $-156\;\mathrm{MeV}$ 
& $-23\;\mathrm{MeV}$ 
& $-94\;\mathrm{MeV}$ 
& $-138.13\pm3.09\;\mathrm{MeV}$
\tabularnewline
\multicolumn{2}{c}{$m_{s}\gamma$} 
& $-107\;\mathrm{MeV}$ 
& $152\;\mathrm{MeV}$ 
& $-53\;\mathrm{MeV}$ 
& $-91.75\pm2.08\;\mathrm{MeV}$
\tabularnewline
\multicolumn{2}{c}{$c_{\overline{10}}$} 
& $0.084$ 
& $0.088$ 
& $0.037$ 
& $0.046\pm0.0002$
\tabularnewline
\hline 
\multicolumn{2}{c}{$N_{\overline{10}}$} 
&  &  &  
& $1701\pm5\;{\rm MeV}$
\tabularnewline
\hline 
\end{tabular}
}
\label{tab:4}
\end{table}

\section{Widths of $\Theta^+$ and $N^*(1685)$}
As discussed by Prasza{\l}owicz, the narrow width is a peculiar but not
unnatural feature of the $\Theta^+$\cite{Praszalowicz:2024zsy}. The
DIANA Collaboration consistently reported a very small value for it:
$\Gamma_{\Theta^+}=0.34\pm 0.10$ MeV\cite{DIANA:2013mhv}, which makes
searching for the $\Theta^+$ tremendously difficult. A key prediction
of Ref.\cite{Diakonov:1997mm} was that the decay width of the
$\Theta^+$ would be rather small.\footnote{Jaffe~\cite{Jaffe:2004qj}
  pointed out a numerical mistake in deriving the decay width of the
  $\Theta^+$ in the paper by Diakonov et
  al.~\cite{Diakonov:1997mm}. If it is corrected, the
  $\Gamma_{\Theta^+}$ would become around 30 MeV. However, as  
  Diakonov et al. refuted in Ref.~\cite{Diakonov:2004ai}, the refined
  analyses on the $\Gamma_{\Theta}$ showed that it is indeed
  small~\cite{Ellis:2004uz, Yang:2013tka}. As 
  will be discussed in this section, more sophisticated analyses within
  the $\chi$QSM   yielded a width of the $\Theta^+$ smaller than 1
  MeV. The smallness of the predicted $\Theta^+$ decay width is a firm
  conclusion from the $\chi$QSM.}, which distinguished it from
previous works in the 1960s and 1970s. In this section, we will
explicitly show that the small decay width of the $\Theta^+$ is
natural in the current approach. 

The collective operator for the axial-vector constant is given as 
\begin{eqnarray}
\hat{g}_A &=&a_1 D_{X3}^{(8)} \;+\; a_2d_{pq3}D_{Xp}^{(8)}\,\hat{J}_q
\;+\; \frac{a_3}{\sqrt{3}} D_{X8}^{(8)}\,\hat{J}_3 \cr 
&+& \frac{a_4}{\sqrt{3}}d_{pq3} D_{Xp}^{(8)}\,D_{8q}^{(8)} \;+ \;a_5\left(
D_{X3}^{(8)}\,D_{88}^{(8)}+D_{X8}^{(8)}\,D_{83}^{(8)}\right)
\cr 
&+& a_6\left(D_{X3}^{(8)}\,D_{88}^{(8)}-D_{X8}^{(8)}\,D_{83}^{(8)}\right),   
  \label{eq:10}
\end{eqnarray}
where $a_1$, $a_2$, and $a_3$ are the SU(3) symmetric terms, which
correspond to $G_0$, $G_1$, and $G_2$~\cite{Diakonov:1997mm,
Praszalowicz:2024zsy}, whereas $a_4$, $a_5$, and $a_6$ arise from
the explicit SU(3) symmetry breaking. These dynamical parameters can
either be determined using experimental data from hyperon
semileptonic decays~\cite{Yang:2015era} or be computed within the
$\chi$QSM~\cite{Ledwig:2008rw}. As shown by Diakonov et al., in the
limit of small soliton size (nonrelativistic limit), the coupling
constant $G_{\Theta N K}$ vanishes~\cite{Diakonov:1997mm,
Praszalowicz:2024zsy}, which explains why the width of the
$\Theta^+$ is naturally small. 

\begin{figure}[ht]
  \centering
\includegraphics[scale=0.6]{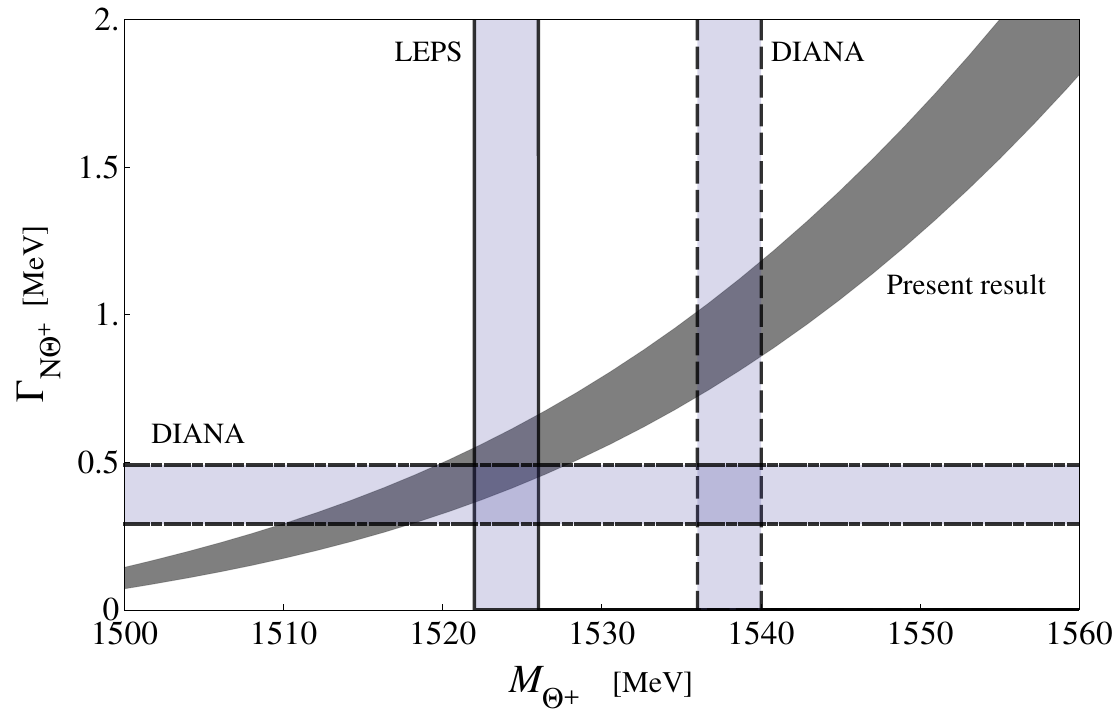}  
  \caption{The dependence of the decay width $\Gamma_{N\Theta^+}$ for
the $\Theta^+\to KN$ decay on $M_{\Theta^+}$. The vertical shaded bars
bounded with the solid and dashed lines denote the measured values of
the $\Theta^+$ mass with uncertainties by the LEPS and DIANA
collaborations, respectively. The horizontal shaded region draws the
values of the $N^*$ mass with uncertainty taken from
Ref.~\cite{Kuznetsov:2008ii}. The sloping shaded 
region represents the present results of the $M_{\Theta^+}$ dependence
of $\Gamma_{N\Theta^+}$.}
  \label{fig:3}
\end{figure}
The decay width of the $\Theta^+$ depends on its
mass. Figure~\ref{fig:3} depicts the mass dependence of the decay
width for $\Theta^+\to KN$. When we use $M_{\Theta^+}=(1524\pm 5)$
MeV, the width of the $\Theta^+$ turns out to be
$\Gamma_{\Theta^+}=(0.5\pm 0.1)$ MeV, which is close to the DIANA
result, $\Gamma_{\Theta^+}=(0.34\pm 0.1)$ MeV. On the other hand, if
we use the mass from the DIANA data, $M_{\Theta^+}=(1538\pm 2)$ MeV,
we obtain $\Gamma_{\Theta^+}\approx 1$ MeV. Note that the prediction
from the $\chi$QSM is $\Gamma_{\Theta^+}=0.71$ MeV, which is also very
small. Thus, both the current approach and the $\chi$QSM yield small
values for the $\Theta^+$ decay width. 

\section{A different summary}
In remembering Maxim Polyakov, we are reminded of a physicist whose 
scientific conviction remained unwavering even in challenging times. The 
initial excitement following his 1997 prediction of the $\Theta^+$ 
pentaquark with Diakonov and Petrov led to numerous experimental searches 
and theoretical works. However, the field reached a critical turning point 
in 2006 when the CLAS Collaboration at Jefferson Lab, using photoproduction 
experiments similar to those that had previously yielded positive results, 
reported null results for the $\Theta^+$. This, combined with null results 
from other high-energy facilities, led to a dramatic decline in light 
pentaquark research. Interestingly, a subsequent reanalysis by Amaryan 
et al., including some members of the CLAS Collaboration, found evidence for 
the $\Theta^+$ in ${}^1$H($\gamma$,$K^0_S$)X via interference with 
$\phi$-meson production, adding another perspective to the ongoing debate. 
The contrast between various experimental results, including the continued 
positive signals from DIANA ($\Gamma_{\Theta^+} = 0.34 \pm 0.10$ MeV) and 
LEPS collaborations, remains an intriguing puzzle in hadron physics. 

The photographs given in Figs.~\ref{fig:1} and \ref{fig:3} included in
this review capture not just the professional collaborations but also
the warm friendships that characterized Maxim's approach to
physics. They remind us of his Siberian hospitality and his belief in
direct, honest scientific discourse - a tradition he traced to the
Landau and Gribov school. His emphasis on physical understanding over
rhetoric and his genuine openness to scientific debate exemplified the
best traditions of theoretical physics. I include three more pictures
in Fig.~\ref{fig:3}, showing the authors of the 1997 paper. 
\begin{figure}
  \centering
 \includegraphics[scale=0.4]{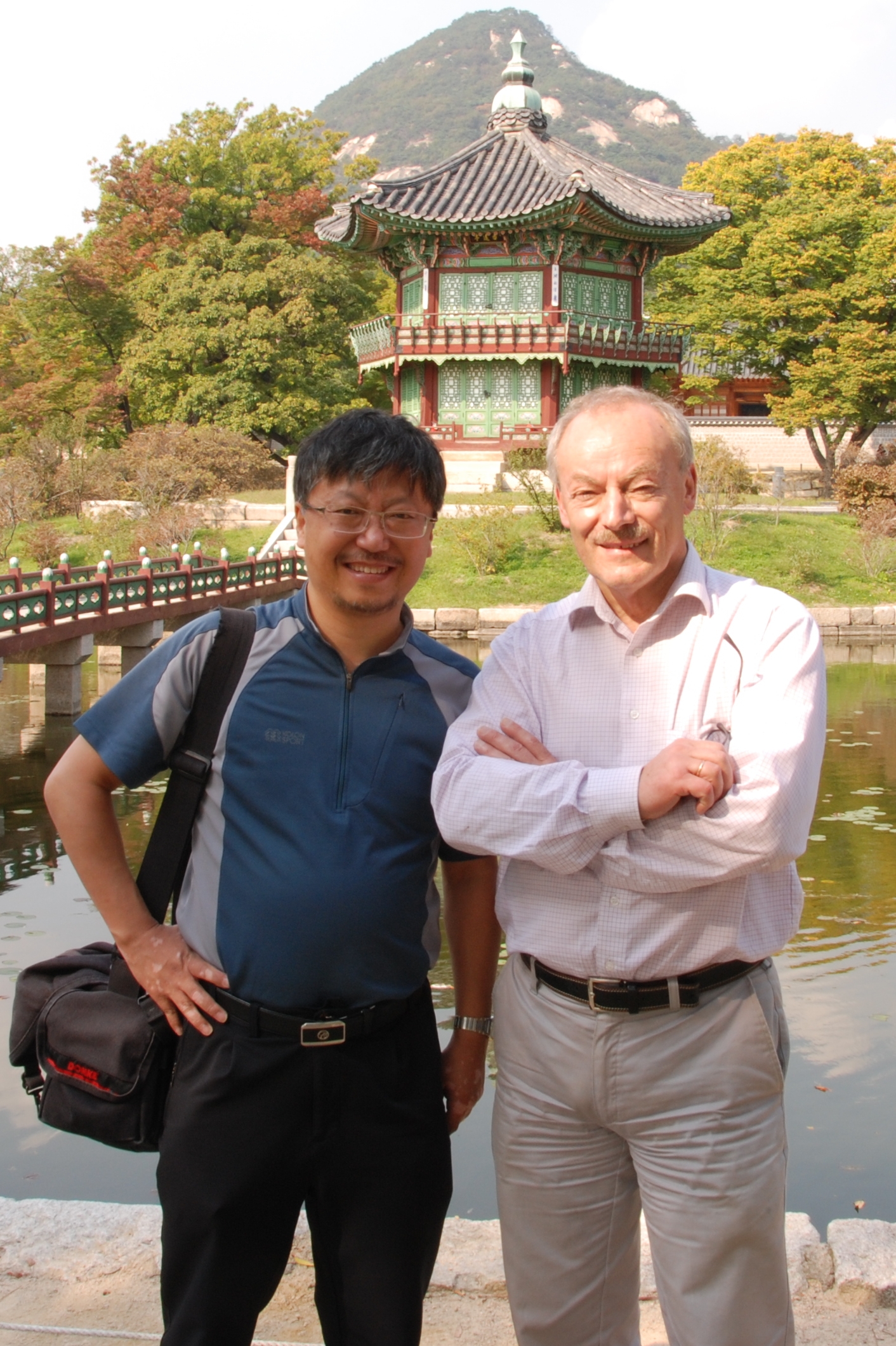}
 \includegraphics[scale=0.2]{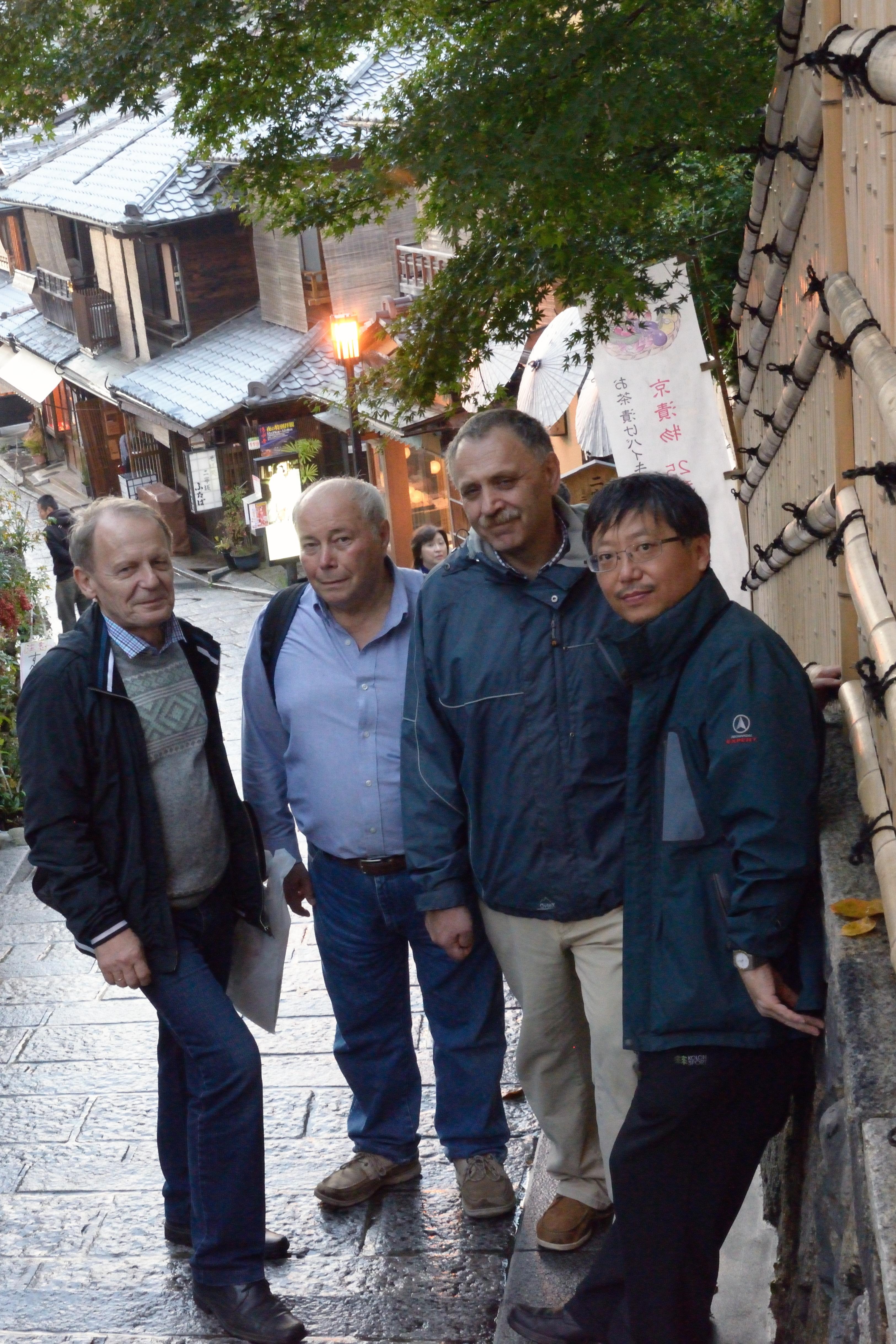}\\
 \includegraphics[scale=0.1765]{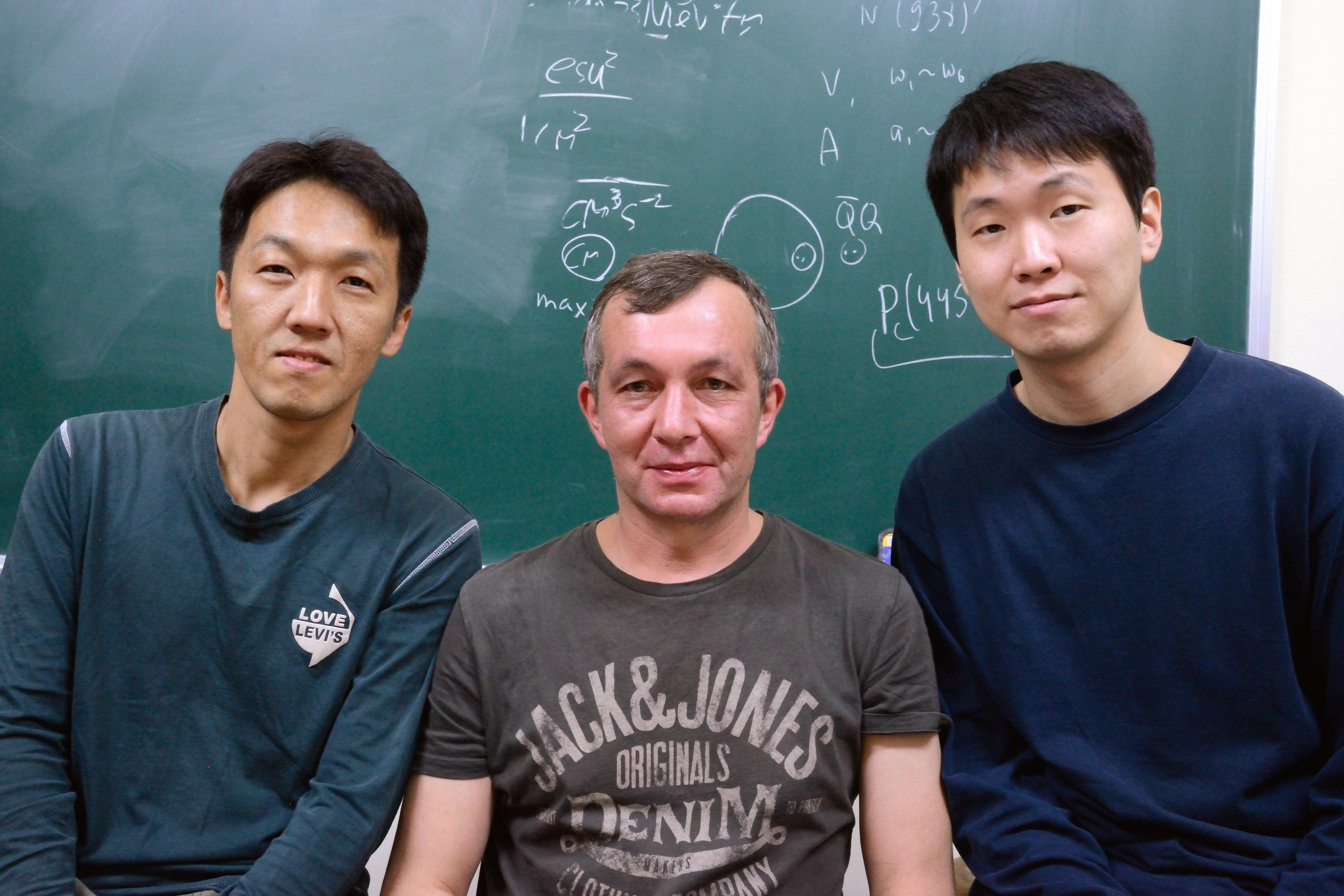} 
  \caption{Upper-left panel: H.-Ch. Kim (left) and Dmitri Diakonov
    (right) in Gyeongbokgung Palace, Seoul. Upper-right panel: Anatoly 
    Gridnev, Igor Strakovsky, Viktor Petrov, and H.-Ch. Kim in Kyoto,
    from the left to the right. Lower panel: Maxim with his former
    Korean students, Ghil-Seok Yang and Hyeon-Dong Son (from the
    left to the right).}   
\label{fig:3}
\end{figure}

As new experimental groups prepare to search for pentaquark states, 
the questions Maxim grappled with remain relevant. His untimely passing 
leaves a void in hadronic physics community, but his work continues 
to illuminate our path forward, particularly as we seek to understand the 
apparent contradictions between different experimental approaches to 
pentaquark searches. His legacy lives on in both the theoretical framework 
he helped develop and the scientific integrity he consistently demonstrated 
throughout his career. 

\section*{Acknowledgments}
I want to express special gratitude to Ghil-Seok Yang, a former
student of Maxim Polyakov. I have worked with him on light and heavy
pentaquarks for 20 years. A major part of the present work stems from
our collaboration. I am also grateful to Igor Strakovsky for giving me
the opportunity to write about Maxim. I am very thankful to Micha{\l}
Prasza{\l}owicz who made invaluable comments on the decay width of the
$\Theta^+$. 
The present work was supported by the Basic Science Research Program
through the National Research Foundation of Korea funded by the Korean
government (Ministry of Education, Science and Technology, MEST),
Grant-No. 2021R1A2C2093368 and 2018R1A5A1025563. 
The author would also like to acknowledge the long-term workshop on
HHIQCD2024 at the Yukawa Institute for Theoretical Physics
(YITP-T-24-02), during which I discussed pentaquark baryons with
several participants, in particular with Atsushi Hosaka.

\end{document}